
\def\J{$J/\psi$}
\def\j{J/\psi}
\def\P{$\psi'$}

\def\U{$\Upsilon$}

\def\c{c{\bar c}}

\def\E{\epsilon}

\def\q{q{\bar q}}
\def\Q{Q{\bar Q}}

\def\e{\epsilon}

\def\lsim{\raise0.3ex\hbox{$<$\kern-0.75em\raise-1.1ex\hbox{$\sim$}}}
\def\gsim{\raise0.3ex\hbox{$>$\kern-0.75em\raise-1.1ex\hbox{$\sim$}}}

\newcount\REFERENCENUMBER\REFERENCENUMBER=0
\def\REF#1{\expandafter\ifx\csname RF#1\endcsname\relax
               \global\advance\REFERENCENUMBER by 1
               \expandafter\xdef\csname RF#1\endcsname
                   {\the\REFERENCENUMBER}\fi}
\def\reftag#1{\expandafter\ifx\csname RF#1\endcsname\relax
               \global\advance\REFERENCENUMBER by 1
               \expandafter\xdef\csname RF#1\endcsname
                      {\the\REFERENCENUMBER}\fi
             \csname RF#1\endcsname\relax}
\def\ref#1{\expandafter\ifx\csname RF#1\endcsname\relax
               \global\advance\REFERENCENUMBER by 1
               \expandafter\xdef\csname RF#1\endcsname
                      {\the\REFERENCENUMBER}\fi
             [\csname RF#1\endcsname]\relax}
\def\refto#1#2{\expandafter\ifx\csname RF#1\endcsname\relax
               \global\advance\REFERENCENUMBER by 1
               \expandafter\xdef\csname RF#1\endcsname
                      {\the\REFERENCENUMBER}\fi
           \expandafter\ifx\csname RF#2\endcsname\relax
               \global\advance\REFERENCENUMBER by 1
               \expandafter\xdef\csname RF#2\endcsname
                      {\the\REFERENCENUMBER}\fi
             [\csname RF#1\endcsname--\csname RF#2\endcsname]\relax}
\def\refs#1#2{\expandafter\ifx\csname RF#1\endcsname\relax
               \global\advance\REFERENCENUMBER by 1
               \expandafter\xdef\csname RF#1\endcsname
                      {\the\REFERENCENUMBER}\fi
           \expandafter\ifx\csname RF#2\endcsname\relax
               \global\advance\REFERENCENUMBER by 1
               \expandafter\xdef\csname RF#2\endcsname
                      {\the\REFERENCENUMBER}\fi
            [\csname RF#1\endcsname,\csname RF#2\endcsname]\relax}
\def\refss#1#2#3{\expandafter\ifx\csname RF#1\endcsname\relax
               \global\advance\REFERENCENUMBER by 1
               \expandafter\xdef\csname RF#1\endcsname
                      {\the\REFERENCENUMBER}\fi
           \expandafter\ifx\csname RF#2\endcsname\relax
               \global\advance\REFERENCENUMBER by 1
               \expandafter\xdef\csname RF#2\endcsname
                      {\the\REFERENCENUMBER}\fi
           \expandafter\ifx\csname RF#3\endcsname\relax
               \global\advance\REFERENCENUMBER by 1
               \expandafter\xdef\csname RF#3\endcsname
                      {\the\REFERENCENUMBER}\fi
[\csname RF#1\endcsname,\csname RF#2\endcsname,\csname
RF#3\endcsname]\relax}
\def\refand#1#2{\expandafter\ifx\csname RF#1\endcsname\relax
               \global\advance\REFERENCENUMBER by 1
               \expandafter\xdef\csname RF#1\endcsname
                      {\the\REFERENCENUMBER}\fi
           \expandafter\ifx\csname RF#2\endcsname\relax
               \global\advance\REFERENCENUMBER by 1
               \expandafter\xdef\csname RF#2\endcsname
                      {\the\REFERENCENUMBER}\fi
            [\csname RF#1\endcsname,\csname RF#2\endcsname]\relax}
\def\Ref#1{\expandafter\ifx\csname RF#1\endcsname\relax
               \global\advance\REFERENCENUMBER by 1
               \expandafter\xdef\csname RF#1\endcsname
                      {\the\REFERENCENUMBER}\fi
             [\csname RF#1\endcsname]\relax}
\def\Refto#1#2{\expandafter\ifx\csname RF#1\endcsname\relax
               \global\advance\REFERENCENUMBER by 1
               \expandafter\xdef\csname RF#1\endcsname
                      {\the\REFERENCENUMBER}\fi
           \expandafter\ifx\csname RF#2\endcsname\relax
               \global\advance\REFERENCENUMBER by 1
               \expandafter\xdef\csname RF#2\endcsname
                      {\the\REFERENCENUMBER}\fi
            [\csname RF#1\endcsname--\csname RF#2\endcsname]\relax}
\def\Refand#1#2{\expandafter\ifx\csname RF#1\endcsname\relax
               \global\advance\REFERENCENUMBER by 1
               \expandafter\xdef\csname RF#1\endcsname
                      {\the\REFERENCENUMBER}\fi
           \expandafter\ifx\csname RF#2\endcsname\relax
               \global\advance\REFERENCENUMBER by 1
               \expandafter\xdef\csname RF#2\endcsname
                      {\the\REFERENCENUMBER}\fi
        [\csname RF#1\endcsname,\csname RF#2\endcsname]\relax}
\def\refadd#1{\expandafter\ifx\csname RF#1\endcsname\relax
               \global\advance\REFERENCENUMBER by 1
               \expandafter\xdef\csname RF#1\endcsname
                      {\the\REFERENCENUMBER}\fi \relax}

%

\def\NP{{ Nucl.\ Phys.\ }}
\def\PL{{ Phys.\ Lett.\ }}
\def\PR{{ Phys.\ Rev.\ }}

\def\PRL{{ Phys.\ Rev.\ Lett.\ }}

\def\ZP{{ Z.\ Phys.\ }}

%
\magnification=1200
\hsize=16.0truecm
\vsize=23.0truecm
\baselineskip=13pt
\pageno=0
\def\la{\Lambda_{\rm QCD}}

\hfill CERN-TH/95-24\par
\hfill BI-TP 95/05\par
\vskip 2 truecm
\centerline{\bf HARD PROBES OF DENSE MATTER$^*$}\footnote{}{
{}~~\par\noindent
Plenary Talk at {\sl Quark Matter '95},
Monterey/California,
January 9 - 13, 1995.\medskip\noindent
CERN-TH/95-24\par\noindent
BI-TP 95/05\par\noindent
February 1995}
\vskip 1.5 truecm
\centerline{\bf Helmut Satz}
\bigskip
\centerline{Theory Division, CERN}
\centerline{CH-1211 Geneva, Switzerland}
\medskip
\centerline{and}
\medskip
\centerline{Fakult\"at f\"ur Physik, Universit\"at Bielefeld}
\centerline{D-33501 Bielefeld, Germany}
\vskip 2 truecm
\centerline{\bf Abstract:}
Direct probes for the QGP must be hard enough to resolve sub-hadronic
scales ($\ll \la^{-1}$) and distinguish confined and deconfined media.
This can be achieved by fast colour charges (jets) and heavy quark
resonances (quarkonia). After a general survey, we study quarkonia as
confinement probe and show in particular that confined matter is
transparent, deconfined matter opaque to \J's.
\par
\vfill\eject
\def\S{${\tilde \sigma}_{\c}$}
\def\s{\tilde{\sigma}_{\c}}
\def\Q{$Q{\bar Q}$}
\def\q{\q{\bar q}}
\def\la{\Lambda_{\rm QCD}}
{}~~\vskip 2 truecm
\noindent
{\bf HARD PROBES OF DENSE MATTER}
\bigskip\bigskip\noindent
Helmut Satz
\bigskip\noindent
Theory Division, CERN, CH-1211 Geneva 23, Switzerland
\par\noindent
and
\par\noindent
Fakult\"at f\"ur Physik, Universit\"at Bielefeld, D-33501 Bielefeld,
Germany
\bigskip\bigskip
Direct probes for the QGP must be hard enough to resolve sub-hadronic
scales ($\ll \la^{-1}$) and distinguish confined and deconfined media.
This can be achieved by fast colour charges (jets) and heavy quark
resonances (quarkonia). After a general survey, we study quarkonia as
confinement probe and show in particular that confined matter is
transparent, deconfined matter opaque to \J's.
\bigskip\medskip
\noindent{\bf 1.\ Probing the Quark-Gluon Plasma}
\medskip
In high energy nuclear collisions, two beams of partons collide; the
partons are initially confined to the colliding nucleons. This
confinement can be verified, e.g.,\ by studying primary high mass
Drell-Yan dilepton production and observing, except for possible
nuclear shadowing effects, the same
parton distribution functions as in deep inelastic lepton-nucleon
collisions. After the primary collision, we expect abundant
multiple interactions, leading to a rapid increase of entropy, quick
thermalisation and hence the production of strongly interacting matter.
The fundamental question is whether confinement survives this
thermalisation. If it does, we have hadronic matter -- if not, a
quark-gluon plasma (QGP). We expect confinement to be lost if the
density of partons sufficiently surpasses that present in a
hadron-hadron interaction, so that partons can no longer be
assigned to specific hadrons. How can we probe if this has happened?
\par
The QGP is a {\sl dense} system of {\sl deconfined} quarks and gluons.
Its density is in fact the reason for deconfinement:
in a sufficiently dense medium, the long-range confining forces
become screened, so that only short-range ($\ll \Lambda_{\rm QCD}^{-1}$)
interactions between quarks and gluons remain. To study such a medium
and determine its nature, we therefore need probes which are {\sl hard}
enough to resolve the short sub-hadronic scales and which can distinguish
between {\sl confined} and {\sl deconfined} quarks and gluons.
In addition, the probe must survive the subsequent evolution of the
medium; therefore it certainly cannot be in equilibrium with the later
stages of matter. Two hard, strongly interacting signals produced before
equilibration and distinct from the medium have been proposed as
{\sl external} probes for confinement/deconfinement:
\item{--}{{\sl hard} quarks or gluons (jets) \refs{Bj}{Miklos}, and}
\item{--}{{\sl heavy} quark-antiquark resonances (charmonium, bottonium)
\refs{Matsui}{Khar3}.}
\par\noindent
Jet and quarkonium production are rather well understood in
hadron-hadron collisions, where they are accounted for in terms of
perturbative QCD and hadronic parton distribution functions
\refs{Jets}{Quarko}.
In both cases, the initially formed state ($Q{\bar Q}$, $q$ or $g$) is
in general coloured, and it has an intrinsic mass or momentum scale
much larger than $\Lambda_{\rm QCD}$. For jets, this is also the state
to be used as probe, since the behaviour of a fast colour charge
passing through confined matter differs from that in a deconfined medium
\ref{Miklos/XNW}. In confined matter, the colour charge loses energy as
it passes from one hadron to the next through the ``interhadronic"
vacuum, and the energy loss is determined by the string tension
$\sigma$ acting on the colour charge as it leaves the field of a hadron
\ref{Khar1}.
In a deconfined medium, the crucial quantity is the colour screening
parameter $\mu$ (the inverse of the screening radius), which determines
with how many other charges the passing colour charge can interact and
thus also how much energy it can lose per unit
length \refs{Gyu-Wa}{Baier}.
\par
For fast quarkonia, the situation is similar; they will pass through
the medium while still in a coloured state \ref{Khar1}, and hence they
can
be used as probe in the same way as jets. In addition, however, we can
consider slow quarkonia, which have become full physical resonances
within a hadronic volume around the \Q~formation point and thus
traverse the medium as colour singlets. Since the intrinsic spatial
scales of \J~and \U, determined by the heavy quark masses and the
binding energies, nevertheless remain much smaller than the hadronic
size $\Lambda_{\rm QCD}^{-1}$, they interact only with the partons
within a big, light hadron and not with the hadron as a whole. They are
thus able to probe the partonic state of any medium. In particular,
they are essentially unaffected by the soft gluons which make up
confined matter, while the hard gluons present in a QGP will dissociate
them \refs{Matsui}{Khar3}.
\par
For both quarkonia and jets, thermal production in the expected
temperature range ($T~\lsim~0.5$ GeV) is excluded by the mass or
momentum scales involved; we can therefore be quite sure that such
signals were produced prior to QGP formation. They will also not
reach an equilibrium with later stages of the medium.
Hard jets and fast quarkonia require too much of an energy loss for
this, while slow quarkonia, as noted, are either dissociated
or not affected by the medium.
\par
For both proposed probes, initial state and/or
pre-equilibrium nuclear effects can occur before QGP formation.
Primary quarks and gluons may undergo multiple scattering or experience
shadowing in the nucleus {\sl before}
they interact to form a \Q~pair or a hard transverse parton. These
effects have to be understood and taken into account before any QGP
analysis \refs{GuptaZP}{Khar2}. It is therefore necessary to
study them in processes which are not effected by the subsequent
medium, such as
the production of hard direct photons \ref{Cley} or of high
mass Drell-Yan dileptons \ref{VesaHP}. In these cases, we have only
annihilation or bremsstrahlung of the incident partons;
the resulting electromagnetic signal leaves the system unaffected by
any subsequent medium and its evolution. If such reactions show nuclear
effects, then these are presumably due to initial state phenomena.
The overall rates for the production of open charm or beauty can
fulfill similar functions \refs{Sridhar}{OpenCharm}.
\par
In addition to the external probes just mentioned, there are
hard electromagnetic signals (photons, dileptons), produced by
annihilation or bremsstrahlung of the constituents of the medium
\refs{Feinberg}{Shuryak}. These will escape unaffected by the subsequent
evolution of the medium and could thus be {\sl internal} probes
of its early state. Thermal dileptons (or photons), if observable,
will certainly povide information about the temperature of the medium
from which they are emitted \refs{Kajantie}{Goldmann}. It is not clear
if they can also tell us something about its nature. In contrast
to the elementary annihilation process of a quark and an antiquark,
the annhilation of two hadrons into a hard dilepton involves more than
one $q{\bar q}$ pair. This introduces a hadronic form factor, which in
turn causes the hadronic annihilation cross section to decrease faster
with dilepton mass $M$ than the $q {\bar q}$ annihilation cross section
(see \ref{Khar4} for a recent discussion). This could in principle make
the dependence of thermal dilepton production on $M$ in a confined
medium different from that in
a deconfined medium. However, it is not clear up to
now if hard thermal dileptons or photons can be experimentally
identified and studied \ref{Vesa}.
\par
In the main part of this survey, we will now consider in detail the
use of quarkonium production as a probe to check if dense strongly
intercting matter is deconfined, i.e., if it forms a QGP. We concentrate
on quarkonium for several reasons.\ \J~suppression was predicted
\ref{Matsui} to be the consequence of QGP formation, and such a
suppression was subsequently indeed observed in high energy
nuclear collisions at the CERN-SPS \ref{NA38}. This
triggered an intensive study of possible alternative origins of such a
suppression. Hence the analysis necessary to establish an
unambiguous probe for deconfinement has been carried much further here
than for jets and can provide a good illustration of what needs to be
done before drawing any conclusions. In particular, we must understand
theoretically and experimentally
\item{--}{the dynamics of the process to be used as probe,}
\item{--}{how the probe reacts to confined matter,}
\item{--}{how it reacts to deconfined matter, and}
\item{--}{how it is influenced by initial state and/or pre-equilibrium
effects.}
\par\noindent
In the following, we shall address mainly the first three points. The
role of initial state and pre-equilibrium nuclear effects has been and
still is under investigation, also for jet production in $p-A$ and $A-B$
collisions.
\medskip
\noindent
{\bf 2.\ Quarkonium Production in Hadron-Hadron Collisions}
\medskip
In this section, we shall sketch the basic dynamics of quarkonium
production in hadron-hadron collisions \ref{Quarko}.
To be specific, we shall speak
about charmonium states; but everything said holds for bottonium as well.
\par
The first stage of charmonium formation is the production of a $\c$
pair; because of the large quark mass, this process can
be described by perturbative QCD (Fig.\ 1). A parton from the
projectile interacts with one from the target; the parton
distributions within the hadrons determined e.g.\ by deep inelastic
lepton-hadron scattering. Initially, the $\c$ pair will in general be in
a colour octet state. It subsequently neutralises its colour and binds
to a physical resonance, such as \J, $\chi_c$ or \P. Colour
neutralisation
occurs by interaction with the surrounding colour field; this and the
subsequent resonance binding are presumably of non-perturbative nature
 (``colour evaporation" \ref{CE}). In the evaporation process,
the $\c$ pair can either combine with light quarks to form open
charm mesons ($D$ and $\bar D$) or bind with each other to form a
charmonium state.
\par
\vskip 6truecm
\centerline{Fig.\ 1: Production of a $c{\bar c}$
pair by gluon fusion (a) and $q{\bar q}$ annihilation (b)}
\medskip
The basic quantity in this description is the total sub-threshold
charm cross section, obtained by integrating the perturbative $\c$
production over the mass interval from $2m_c$ to $2m_D$. At high
energy, the dominant part of \S~comes from gluon fusion (Fig.\ 1a), so
that we have
$$
\s(s) = \int_{2m_c}^{2m_D} d\hat s \int dx_1 dx_2~g(x_1)~g(x_2)~
\sigma(\hat s)~\delta(\hat s-x_1x_2s), \eqno(1)
$$
with $g(x)$ denoting the gluon density and
and $\sigma$ the $gg \to \c$ cross section. In pion-nucleon collisions,
there are also significant quark-antiquark contributions (Fig.\ 1b),
which become dominant at low energies.
The essential prediction of the colour evaporation model is
that the production cross section of any charmonium state $i$
is given by
$$
\sigma_i(s)~=~f_i~\s(s), \eqno(2)
$$
where $f_i$ is a constant which for the time being
has to be determined empirically. In other words, the energy
dependence of any charmonium production cross section is predicted to be
that of the perturbatively calculated sub-threshold charm cross section.
As a consequence, the production ratios of different charmonium
states
$$
{\sigma_i(s)\over \sigma_j(s)} = {f_i\over f_j} = {\rm const.} \eqno(3)
$$
are predicted to be energy-independent.
\par
These predictions have recently been compared \ref{Quarko}
to the available data. In Figs.\ 2 and 3,
we see that the energy-dependence is
well described for both \J~and \U~production. We note in particular that
for \U~production this holds up to 1.8 TeV, so that these rates can be
given with considerable confidence for RHIC and LHC energies. In Figs.\
4 and 5, also the predicted energy-independence of production ratios is
found to hold, again up to Tevatron energy. Here it should be noted that
the CDF data for the ratio \P/(\J) are taken at large transverse momenta
($5\leq p_T\leq 15$ GeV), while the lower energy data are integrated
over $p_T$, with the low $p_T$ region dominant. Hence colour evaporation
appears to proceed in the same way at both small and large $p_T$. An
understanding of this and the theoretical determination of
the coefficients $f_i$ in eq.\ (2) are presently the great challenges in
the study of quarkonium production.
\medskip\noindent
{\bf 3.\ Quarkonium Production in Hadron-Nucleus Collisions}
\medskip
Given the theory of quarkonium production in hadron-hadron
collisions, we want to use hadron-nucleus collisions to study the
effect of a confined medium on the proposed probe. Before doing this,
however, we have to return to the evolution of quarkonium production
already mentioned above. Experimentally, quarkonium production
is studied through its decay dileptons (generally
dimuons); hence these have to be sufficiently hard to pass through the
absorber of a muon spectrometer. As a consequence, the quarkonia
measured so far in fixed nuclear target $p-A$ collisions are always
fast in the nuclear rest frame. This means that they traverse the
nucleus in
the very early stage of their evolution, in which the \Q~pair is still
in a colour octet state. To confirm this experimentally, we compare
\J~and \P~production; the fully formed resonances differ in size by a
factor four, so that their interaction in a nucleus should be quite
different. As nascent small colour singlets, they would suffer
little or no interaction, because of colour transparency. Data
\ref{E772},
however, show that there is a significant suppression of fast quarkonium
production in nuclei, compared to that on nucleons, but that this
suppression is the same for \P~and \J. The production ratio of the two
states is moreover found to be $A$-independent and the same as
in hadron-hadron collisions (Fig.\ 6), indicating that the colour
evaporation occurs well outside the nucleus.
What such experiments study is therefore the
passage of a colour charge through (confined) nuclear matter. This is
also studied in jet production, and it would
certainly be interesting to compare in detail jet and quarkonium data
from $p-A$ collisions.
The passage of fast charges through matter is a very fascinating topic;
one of the most striking theoretical conclusions, the suppression of
soft radiation -- the so-called Landau-Pomeranchuk effect \ref{Mikael}
-- has just recently been confirmed experimentally in high energy
experiments \ref{SLAC}. It is due to quantum-mechanical coherence
effects, and these can occur as well in the case of colour charges
\ref{Khar2}. In a confined QCD medium,
the colour charge will in addition to such ``nuclear shadowing" suffer
an energy loss on its way ``out", and this will shift both
quarkonium and jet production in nuclei to lower momenta
\ref{Khar1}.
\par
We want to concentrate here on the use of fully formed
quarkonium states as probes. For $p-A$ collisions, this requires the
study of slow quarkonia in the nuclear rest frame and hence seems
experimentally very difficult, since slow charmonium decay dimuons will
not make it through the absorber. A rather novel solution
to this problem is offered by the advent of heavy ion {\sl beams}
\ref{Khar3}. Fast quarkonia produced by a lead beam incident
on a hydrogen target lead to fast decay dimuons in the lab system,
so these would be readily detectible. Sufficiently fast quarkonia,
on the other hand, would now be slow in the nuclear (beam) rest frame,
and so provide information about their fate in nuclear matter.
The kinematic region of interest for this consists of quarkonia with
``thermal" momenta, i.e., some 1 - 3 GeV; this is attained in
the forward region, with rapidities $y \geq 5$ (or $0.4 \leq x_F \leq
1.0$) for $A-p$
collisions. Any \J's measured in such an experiment would have been
fully formed before leaving the interaction zone of a single nucleon; it
would thus indeed probe confined matter.
\par
\vskip 8 truecm
\centerline{Fig.\ 6: The $A$-dependence of the ratio $\psi'/\j$}
\bigskip
\noindent
{\bf 4.\ Quarkonium Production as Deconfinement Probe}
\medskip
The ultimate constituents of matter are evidently always quarks and
gluons. What we want to know is if these quarks and gluons are
confined to hadrons or not. As prototype for matter in a confined
state, let us consider an ideal gas of pions. Their momentum distribution
is thermal, i.e., for temperatures not too low it is given by
${\rm exp}(-E_{\pi}/T) \simeq {\rm exp}(-p_{\pi}/T)$. Hence the average
momentum of a pion in this medium is $\langle p_{\pi} \rangle = 3~T$. The
distribution of quarks and gluons within a pion is known from structure
function studies; the gluon density is $g(x) \simeq 0.5 (1-x)^3$ for
large $x=p_g/p_{\pi}$.\footnote{$^{1)}$}{For very small $x$, recent
results
from HERA indicate a steeper increase; however, this does not affect our
considerations here.} As a consequence, the average momentum of a gluon
in confined matter is given by
$$
\langle p_g \rangle_{\rm conf} = {1 \over 5} \langle p_{\pi} \rangle =
{3 \over 5} T. \eqno(4)
$$
Hence in a medium of temperature $T \simeq 0.2$ GeV, the average gluon
momentum is around 0.1 GeV.
In contrast, the distribution of gluons in a deconfined medium is
directly thermal, i.e., ${\rm exp}(-p_g/T)$, so that
$$
\langle p_g \rangle_{\rm deconf} = 3T. \eqno(5)
$$
Hence the average momentum of a gluon in a deconfined medium is five
times higher than in a confined medium\footnote{$^{2)}$}{We could
equally well
assume matter at a fixed energy density, instead of temperature. This
would lead to gluons which in case of deconfinement are approximately
three times harder than for confinement.}; for $T=0.2$ GeV, it
becomes 0.6 GeV.
We thus have to find a way to look for such a hardening of the gluon
distribution in deconfined matter.
\par
The lowest charmonium state \J~provides an ideal probe for
this. It is very small, with a radius $r_{\psi}\simeq 0.2$ fm $\ll
\Lambda^{-1}_{QCD}$, so that \J~interactions with the conventional light
quark hadrons probe the short distance features, the parton
infra-structure, of the latter. It is very strongly bound, with a
binding energy $\epsilon_{\psi} \simeq 0.65$ GeV $\gg \Lambda_{QCD}$;
hence it
can be broken up only by hard partons. Since it shares no quarks or
antiquarks with pions or nucleons, the dominant perturbative interaction
for such a break-up is the exchange of a hard gluon.
\par
We thus see qualitatively how a deconfinement test can be carried out.
If we put a \J~into matter of temperature $T=0.2$ GeV, then
\item{--}{if the matter is confined, $\langle p_g \rangle_{\rm conf}
\simeq
0.1$ GeV, which is too soft to resolve the \J~as a
$\c$ bound state and less than the binding energy $\epsilon_{\psi}$,
so that the \J~survives;}
\item{--}{if the matter is deconfined, $\langle p_g \rangle_{\rm decon}
\simeq$
0.6 GeV, which (with some spread in the momentum distribution) is hard
enough to resolve the \J~and to break the binding, so that the \J~will
disappear.}
\par\noindent
The latter part of our result is in accord with the mentioned prediction
that the formation of a QGP should lead to \J~suppression \ref{Matsui}.
There it was argued that colour screening would prevent any
resonance binding between the perturbatively produced $c$ and ${\bar
c}$, allowing the heavy quarks to separate. At the hadronisation point
of the medium, they would then be too far apart to bind to a \J~and
would therefore form a $D$ and a $\bar D$. Although the details of such
a picture agreed well with the observed \J~suppression \ref{GuptaPL}, it
seemed possible to obtain a similar suppression by absorption in a
purely hadronic medium \ref{Blaizot}, through by collisions of the
type
$$
\j + h \to D + {\bar D} + X. \eqno(6)
$$
Taking into account the partonic substructure of such hadronic
break-up processes, we now see that this is in fact not possible for
hadrons
of reasonable thermal momentum. Our picture thus not only provides a
dynamical basis for \J~suppression
by colour screening, but it also indicates that in fact \J~suppression
in dense matter will occur {\sl if and only if} there is deconfinement.
\par
\refadd{Peskin}
\refadd{Bhanot}
\refadd{Vain}
\refadd{Kaidalov}
\refadd{Khar3}
The theoretical basis of these arguments is provided by the calculation
of the inelastic \J-hadron cross section, $\sigma_{\psi h}$, which
we can then use to check if a \J~can be broken up on its passage
through hadronic matter. Because of the small radius
and large binding energy of the \J, $\sigma_{\psi h}$ can be
calculated in the short-distance analysis of QCD
\refto{Peskin}{Khar3}. The crucial feature for this calculation is
the fact that heavy and tightly bound quarkonium states can be
broken up by scattering on usual light hadrons only through the
exchange of sufficiently hard gluons. The cross section for this is
given by \ref{Bhanot}
$$
\sigma_{g-\psi}(p_g)~\simeq~65~{\rm mb}~\left[{p_g\over E_{\psi}} -
1 \right]^{3/2}\left[{E_{\psi}\over p_g}\right]^5
\Theta(p_g - E_{\psi}),  \eqno(7)
$$
where $E_{\psi}=2M_D - M_{\psi} \simeq 0.64$ GeV is the \J~binding
energy and $p_g$ the momentum of the gluon. The behaviour of
$\sigma_{g-\psi}$ as function of the gluon momentum is shown in Fig.\
7; it is seen that gluons in the momentum range from 0.8 to 2.0 GeV are
most effective in breaking up \J's.
\par
The momentum distribution of gluons within a light hadron of momentum
$p_h$ incident on a \J, $g(x)$, with $x=p_g/p_h$, is a
universal non-perturbative input, determined e.g. from parton counting
rules or from deep inelastic processes. For large $x$ it has
the form $g(x)\simeq (1-x)^k$, with $k\simeq 3$ for mesons, as used
above, and $k\simeq 4$ for nucleons. The resulting inelastic $\j-h$
cross section then becomes \ref{Khar3}
$$
\sigma_{h-\psi}(s) \simeq 3~{\rm
mb}~\left[1-{\lambda_o\over E_h}\right]^{k+5/2} \Theta(E_h - \lambda_o),
\eqno(8)
$$
with $E_h=(m_h^2+p_h^2)^{1/2}$ and $\lambda_o \simeq
(\E_{\psi}+m_h)$.
For low collision energies $\sqrt s$, the cross section is thus
determined by the behaviour of the gluon distribution at large $x$, as
already noted above, and this leads to a very slow growth from
threshold towards the asymptotic value of 3 mb. The functional form of
this behaviour is also shown in Fig.\ 7 for $\pi-\psi$ interactions;
it is essentially unchanged for $\rho-\psi$ or $A-\psi$ collisions,
since it is the gluon distribution in the meson, not the meson mass,
which matters.
\par
The short-distance calculation of the quarkonium-hadron cross section
becomes exact in the limit of large quark mass. Is the charm quark
mass sufficiently large to apply the results of heavy quark theory?
Let us first consider this question theoretically and check possible
non-perturbative contributions to the break-up process in the threshold
region \ref{Larry}. These can be pictured most simply as a quark
rearrangement. Consider putting a \J~``into" a stationary light hadron;
the quarks could then just rearrange their binding pattern
to give rise to transitions such as $\j + N \to \Lambda_c+{\bar D}$ or
$\j+\rho \to D + {\bar D}$. The probability for such a
rearrangement transition can be written as
$$
P_{ra} \sim \int d^3 r~R_h(r)~|\phi_{\psi}(r)|^2, \eqno(9)
$$
where the spatial distribution of the $\c$ bound state is
given by the squared wave function $|\phi_{\psi}(r)|^2$. The function
$R_h(r)$ in eq.\ (4) describes the resolution power of the colour
field inside the light hadron $h$. The average wave length of this
colour field is of order $\la^{-1}$,
and so it cannot resolve the charge content of a very much smaller bound
state; in other words, it doesn't ``see" the heavy quarks in a
bound state of radius $r_{\psi} \ll \la^{-1}$ and hence can't rearrange
bonds. The resolution $R(r)$ approaches unity for $r \simeq
\la^{-1}$; for $r < \la^{-1}$, it rapidly decreases as
$$
R_h(r) \simeq (r\la)^n, ~~r\la < 1,   \eqno(10)
$$
with $n=2$ \ref{Low} or 3 \ref{Peskin}. Note that $R_h(r)$ describes the
colour field of the light hadron and is independent of the heavy quark
state. However, the quarkonium radius decreases with increasing quark
mass, and hence the same holds for the overlap of $R_h(r)$ and the
squared quarkonium wave function: the light hadron quarks can no longer
resolve the small heavy quark bound state.
\par
Quantitatively, this problem can be solved by calculating the tunnelling
rate of the charm quarks inside the \J~from $r=r_{\psi}$ out to a
distance $r\sim \la^{-1}$ at which the light quarks can resolve them.
Such tunneling processes are truly non-perturbative: they cover a
large space-time region, of linear size $\Lambda_{QCD}^{-1}$, and do
not involve any hard interactions. Here the resulting tunnelling rate
is found to be \ref{Larry}
$$
R_{\rm tun} = E_{\psi} \ {\rm exp}\{- 1.2 \sqrt{m_c\ E_{\psi}}/ \la\},
\eqno(11)
$$
where $E_{\psi}\simeq 0.64$ GeV again denotes the \J~binding
energy. With $m_c=1.5$ GeV and $\la=0.3$ GeV, this yields
$R_{\rm tun} \simeq 9.0 \times 10^{-3}~{\rm fm}^{-1}$
for the rate of \J~dissociation by tunnelling. From it, we get
$$
S_{\rm tun} = {\rm exp}\left\{
-\int_0^{t_{\rm max}} dt~R_{\rm tun}\right\}, \eqno(12)
$$
for the \J~survival probability. The maximum time $t_{\rm max}$ which
the \J~can spend adjacent to the light hadron is by
the uncertainty relation about 4 - 5 fm. The resulting
survival probability remains more than 95 \%, so that
there is effectively no \J~break-up by non-perturbative hadron
interactions either.
\par
As emphasized, the inelastic \J-hadron cross section calculated in
short-distance
QCD implies that a \J~of ``thermal" momentum (1 - 3 GeV) cannot be
dissociated in a confined medium. The direct experimental test for this
prediction was already indicated in section 3: it requires the study
of slow \J's in nuclear matter, which is now possible with a
nuclear beam incident on a hydrogen (or other light) target. Such
experiment is evidently of great importance to have a clear experimental
basis for the distinction of \J~production in confined and deconfined
matter.
\par
In the meantime, the approach can be checked on similar
processes. Using the same short-distance analysis, one can
calculate the cross section for the photoproduction of open charm,
$\sigma_{\gamma h \to \c}(s)$. For this reaction, there are data
\ref{data}, and they agree well with the prediction of the heavy quark
analysis \ref{Khar3}.
\par
{}From short-distance QCD we have thus learned that confined matter
cannot dissociate \J's. In deconfined matter this is possible, however:
the cross section for the dissociation of a \J~by gluon collisions was
already given above (eq.\ (7)) and found to be most effective for
momenta around 1 - 2 GeV, i.e., for momenta in the ``thermal" range.
To compare the two cases directly, we average the corresponding
cross sections with a Boltzmann distribution for gluons or pions,
$$
\sigma(T) = \int d^3p~{\rm e}^{-|p|/T} \sigma(p)~\Big/~
\int d^3p~{\rm e}^{-|p|/T} .  \eqno(13)
$$
The result is shown a in Fig.\ 8. For gluon-dissociation, the peak
occurs for temperatures around 300 - 400 MeV; for hadrons, the
cross section becomes comparable only for $T \simeq 1$ GeV or higher.
\par
In closing this section, we comment briefly on the
experimental consequences for charmonium production
in nucleus-nucleus collisions. There are two ways to probe:
\item{--}{compare charmonium production to an unaffected signal, e.g.,
study \J~production with respect to the Drell-Yan continuum, or}
\item{--}{compare different charmonium states with each other, e.g.,
study the ratio of \P/(\J) production.}
\par\noindent
In both cases, the behaviour of the relevant ratio as function of the
initial energy density $\e_0$ is of particular interest. As
illustration, we consider the $\e_0$ dependence of \J~production. The
observed
\J's are in part ($\sim$40\%) due to $\chi$-decay ($\chi \to
\j+\gamma$)
\ref{Lemoigne}. The larger and more loosely bound
$\chi$'s can be broken up by softer gluons than
needed for the dissociation of ground state \J's, so that they could
also be suppressed in confined matter. Confinement thus allows (at
most) a suppression of the $\chi$ component of the \J's and hence
requires
a saturation of the suppression with increasing energy density, at a
survival probability of about 60\%. In deconfined matter, on the other
hand, also the ground state \J~can be broken up, leading to complete
suppression at large $\e_0$.
\medskip\noindent
{\bf 5. Summary}
\medskip
To study the dense primordial QGP, we need probes which are sufficiently
hard to resolve subhadronic scales ($\ll \la^{-1}$) and which can
distinguish confined and deconfined media. Suitable probes can be either
colourless
(fully formed quarkonia) or coloured (fast \Q~pairs or hard jets). For
\J's, confined matter is transparent, while deconfined matter dissolves
them. The situation here is thus particularly clear: even if the medium
undergoes the quark-hadron transition during the passage of the \J,
the only effect will come from the QGP phase. For jets or fast
colour-octet \Q~pairs, the energy loss in deconfined matter differs
from that in a deconfined medium \refs{Bj}{Miklos};
interesting recent
work \refs{Gyu-Wa}{Baier} may help to turn this difference
into a viable probe.
\vfill\eject
\noindent{\bf Acknowledgements}
\medskip
It is a pleasure to thank the members of the {\sl Hard Probe
Collaboration} for stimu-\break
lating discussions; special thanks
for much help and support go to D. Kharzeev, G. Schuler and R. Vogt.
\bigskip
\noindent
{\bf References}
\medskip
\parindent=13pt
\item{\reftag{Bj})}{J. D. Bjorken, ``Energy Loss of Energetic
Partons in Quark-Gluon Plasma", Fermilab-Pub-82/59-THY, August 1982
(unpublished), and Erratum.}
\par
\item{\reftag{Miklos})}{M. Gyulassy and M. Pl\"umer, \PL B 243
(1990) 432.}
\par
\item{\reftag{Matsui})}{T. Matsui and H. Satz, \PL B 178 (1986)
416.}
\par
\item{\reftag{Khar3})}{D. Kharzeev and H. Satz, \PL B 334 (1994) 155.}
\par
\item{\reftag{Jets})}{K. Eskola and X.-N. Wang, ``High $p_T$ Jet
Production in $p-p$ Collisions", in {\sl
Hard Processes in Hadronic Interactions}, H. Satz and X.-N. Wang (Eds.)}
\par
\item{\reftag{Quarko})}{R. V. Gavai et al., ``Quarkonium
Production in Hadronic Collisions", in {\sl
Hard Processes in Hadronic Interactions}, H. Satz and X.-N. Wang
(Eds.); CERN Preprint CERN-TH.7526/94 (December 1994).}
\par
\item{\reftag{Miklos/XNW})}{M. Gyulassy et al., \NP A 538 (1992) 37c.}
\par
\item{\reftag{Khar1})}{D. Kharzeev and H. Satz, \ZP C 60 (1993) 389.}
\par
\item{\reftag{Gyu-Wa})}{M. Gyulassy and X.-N. Wang, \NP B 420 (1994)
583.}
\par
\item{\reftag{Baier})}{R. Baier, Yu. Dokshitser, S. Peigne and D. Schiff,
``Induced Gluon Radiation in
a Deconfined Medium", Bielefeld Preprint BI-TP-94-57, November 1994.}
\par
\item{\reftag{GuptaZP})}{S. Gupta and H. Satz, \ZP C 55 (1992) 391.}
\par
\item{\reftag{Khar2})}{D. Kharzeev and H. Satz, \PL B 327 (1994) 361.}
\par
\item{\reftag{Cley})}{J. Cleymans et al., ``Prompt Photon
Production in $p-p$ Collisions", in {\sl
Hard Processes in Hadronic Interactions}, H. Satz and X.-N. Wang (Eds.)}
\par
\item{\reftag{VesaHP})}{S. Gavin et al., ``Production of Drell-Yan
Pairs in High Energy Nucleon-Nucleon Collisions", in {\sl
Hard Processes in Hadronic Interactions}, H. Satz and X.-N. Wang (Eds.)}
\par
\item{\reftag{Sridhar})}{H. Satz and K. Sridhar, \PR D 50 (1994)
3557.}
\par
\item{\reftag{OpenCharm})}{R. V. Gavai et al., ``Heavy Quark
Production in $p-p$ Collisions", in {\sl
Hard Processes in Hadronic Interactions}, H. Satz and X.-N. Wang (Eds.)}
\par
\item{\reftag{Feinberg})}{E. L. Feinberg, Nuovo Cim. 43A (1976) 39.}
\par
\item{\reftag{Shuryak})}{E. Shuryak, \PL 78B (1978) 150.}
\par
\item{\reftag{Kajantie})}{K. Kajantie and H. I. Miettinen, \ZP C 9
(1981) 341 and C 14 (1982) 357.}
\par
\item{\reftag{Goldmann})}{G. Domokos and  J. Goldmann, \PR D 23
(1981) 203 and D 28 (1983) 123.}
\par
\item{\reftag{Khar4})}{D. Kharzeev and H. Satz, \PL B 340 (1994) 167.}
\par
\item{\reftag{Vesa})}{See, e.g., P. V. Ruuskanen,
\NP A544 (1992) 169c;\hfill\break
J. I. Kapusta, \NP A566 (1994) 45c.}
\par
\item{\reftag{NA38})}{C. Baglin et al., \PL B220 (1989) 471; B251
(1990) 465, 472; B225 (1991) 459.}
\vfill\eject
\item{\reftag{CE})}{M. B. Einhorn and S. D. Ellis, \PR D12 (1975)
2007;\hfill\par
H. Fritzsch, \PL 67B (1977) 217;\hfill\par
M. Gl\"uck, J. F. Owens and E. Reya, \PR D17 (1978) 2324; \hfill\par
J. Babcock, D. Sivers and S. Wolfram, \PR D18 (1978) 162.}
\par
\item{\reftag{E772})}{D. M. Alde et al., \PRL 66 (1991) 133 and 2285.}
\par
\item{\reftag{Mikael})}{L. D. Landau and I. Ya. Pomeranchuk,
Dokl. Akad. Nauk SSSR 92 (1953) 535 and 735;\hfill\par
E. L. Feinberg and I. Ya. Pomeranchuk,
Suppl.\ Nuovo Cim.\ III, Ser.\ X, No.\ 4 (1956)
652;\hfill\par
M. L. Ter-Mikaelyan, J.E.T.P. 25 (1954) 289 and 296; \hfill \par
A. B. Migdal, \PR 103 (1956) 1811.}
\par
\item{\reftag{SLAC})}{S. R. Klein et al., ``A Measurement of the LPM
Effect", SLAC-PUB-6378 (November 1993).}
\par
\item{\reftag{GuptaPL})}{S. Gupta and H. Satz, \PL B 283 (1992) 439.}
\par
\item{\reftag{Blaizot})}{See e.g., J.-P. Blaizot and J.-Y. Ollitrault,
in {\sl Quark-Gluon Plasma}, R. C. Hwa (Ed.), World Scientific,
Singapore 1990.})
\par
\item{\reftag{Peskin})}{M. E. Peskin, \NP B 156 (1979) 365.}
\par
\item{\reftag{Bhanot})}{G. Bhanot and M. E. Peskin, \NP B 156 (1979)
391.}
\par
\item{\reftag{Vain})}{M. A. Shifman, A. I. Vainshtein and V. I.
Zakharov, \PL 65B (1976) 255.\hfill \break
V. A. Novikov, M. A. Shifman, A. I. Vainshtein and V. I. Zakharov,
\NP B136 (1978) 125.}
\par
\item{\reftag{Kaidalov})}{A. Kaidalov, in {\sl QCD and High Energy
Hadronic Interactions}, J. Tran Thanh Van (Ed.), Ed. Frontieres,
Gif-sur-Yvette 1993.}
\par
\item{\reftag{Larry})}{D. Kharzeev, L. McLerran and H. Satz,
``Non-Perturbative Quarkonium Dissociation in Hadronic Matter",
CERN Preprint CERN-TH/95-27, February 1995.}
\par
\item{\reftag{Low})}{F. E. Low, \PR D12 (1975) 163;\hfill\break
                    S. Nussinov, \PRL 34 (1975) 1268;\hfill\break
J. F. Gunion and H. Soper, \PR D15 (1977) 2617.}
\par
\item{\reftag{data})}{S. D. Holmes, W. Lee and J. E. Wiss, Ann. Rev.
Nucl. Sci. 35 (1985) 397;\hfill \break
T. Nash, in {\sl Proceedings of the International Lepton/Photon
Symposium}, Cornell 1983, p. 329; \hfill\break
L. Rossi, ``Heavy Quark Photoproduction", Genova Preprint
INFN/AE-91-16;\hfill \break
A. Ali, ``Heavy Quark Physics in Photo- and Leptoproduction Processes at
HERA and Lower Energies", DESY Preprint 93-105, 1993.}
\par
\item{\reftag{Lemoigne})}{Y. Lemoigne et al., \PL B 113 (1982) 509;
\hfill\break
L. Antoniazzi et al., \PRL 70 (1993) 383.}
\vfill\bye